\documentclass{aa}
\usepackage{graphics}
 \def\dv{$<V_{abs}-V_{emi}>$}
\begin{document}
\thesaurus{11(11.04.1 11.11.1)}
\title{The ESO Slice Project (ESP) galaxy redshift survey
\thanks{based on observations collected at the European Southern
Observatory, La Silla, Chile.} }
\subtitle{IV. A discussion of systematic biases in galaxy redshift 
determinations}
\author{
A.~Cappi$^1$, 
G.~Zamorani$^{1,2}$, 
E.~Zucca$^{1,2}$, 
G.~Vettolani$^2$, 
R.~Merighi$^1$, 
M.~Mignoli$^1$, 
G.M.~Stirpe$^1$,
C.~Collins$^3$,  
L.~Guzzo$^4$, 
G.~Chincarini$^{4,5}$, 
D.~Maccagni$^6$,
C.~Balkowski$^7$,
V.~Cayatte$^7$, 
S.~Maurogordato$^{7,8}$,
D.~Proust$^7$, 
S.~Bardelli$^9$, 
M.~Ramella$^9$,
R.~Scaramella$^{10}$, 
A. Blanchard$^{11}$,
H.~MacGillivray$^{12}$
}
\institute{ 
Osservatorio Astronomico di Bologna, 
via Zamboni 33, 40126 Bologna, Italy
\and
Istituto di Radioastronomia del CNR, 
via Gobetti 101, 40129 Bologna, Italy
\and
Astrophysics Research Institute, Liverpool John--Moores University, 
Byrom Street, Liverpool L3 3AF, United Kingdom
\and
Osservatorio Astronomico di Brera, 
via Bianchi 46, 22055 Merate (LC), Italy
\and
Universit\`a degli Studi di Milano, 
via Celoria 16, 20133 Milano, Italy
\and
Istituto di Fisica Cosmica e Tecnologie Relative, 
via Bassini 15, 20133 Milano, Italy
\and
Observatoire de Paris, DAEC, Unit\'e associ\'ee au CNRS, D0173 et \`a
l'Universit\'e Paris 7, 5 Place J.Janssen, 92195 Meudon, France
\and
CERGA, Observatoire de la C\^ote d'Azur, 
B.P. 229, 06304 Nice Cedex 4, France
\and
Osservatorio Astronomico di Trieste, 
via Tiepolo 11, 34131 Trieste, Italy
\and
Osservatorio Astronomico di Roma, 
via Osservatorio 2, 00040 Monteporzio Catone (RM), Italy
\and
Universit\'e L. Pasteur, Observatoire Astronomique, 
11 rue de l'Universit\'e, 67000 Strasbourg, France
\and
Royal Observatory Edinburgh, 
Blackford Hill, Edinburgh EH9 3HJ, United Kingdom
}
\offprints{Alberto Cappi}
\mail{cappi@astbo3.bo.astro.it}
\date{Received ~ / Accepted ~}
\titlerunning{ESP IV: biases in redshift determinations}
\authorrunning{A. Cappi et al.}
\maketitle
\begin{abstract}
We present a detailed discussion of the redshift errors associated to the ESO
Slice Project measurements. For a subsample of 742 galaxies
with redshifts determined both from the absorption lines ($V_{abs}$)
and from the emission lines ($V_{emi}$),
we find an average difference $<V_{abs} - V_{emi}> \simeq +100$ km/s. 
We find that a similar effect is present in another, deeper
redshift survey, the Durham/Anglo--Australian Telescope 
faint galaxy redshift survey (Broadhurst et al. 1988), while is absent 
in surveys at brighter magnitude limits.
We have investigated in detail many possible 
sources of such a discrepancy, and we can exclude possible zero--point shifts
or calibration problems.
We have detected and measured systematic velocity differences
produced by the different templates used in the cross--correlation.
We conclude that such differences can in principle explain the effect,
but in this case the non--trivial implication would be that
the best--fitting template does not necessarily give the best velocity 
estimate.
As we do not have any {\em a priori} reason to select a 
template different from the best--fitting one, we did not apply
any correction to the ESO Slice Project velocities.
However, as for a small number of galaxies the effect is so large that 
it is likely to have a physical explanation, we have also taken into account
the possibility that the discrepancy can be partly real: in this case,
it might help to understand the role of gas
outflows in the process of galaxy evolution.
In view of the future large spectroscopic surveys, we stress the importance
of using different templates and making them publicly available, 
in order to assess the amplitude of systematic effects, and to allow a direct
comparison of different catalogues.
\end{abstract}
\begin{keywords}
Galaxies: distances and redshifts; kinematics and dynamics
\end{keywords}
\section{Introduction}

The understanding of the formation, evolution and
present properties of the large--scale structure of the Universe
is a key problem in modern cosmology (see Peebles 1980, 1993).
One of the most important results of the first redshift surveys
was the previously unexpected existence of coherent structures and 
voids at very large scales. 
Explaining these structures was a challenge for popular models of
galaxy formation, but at the same time represented a problem
for the interpretation of results obtained on
small volumes which could not be representative of the Universe.
Therefore, the need of a ``fair sample" 
of the Universe, in order to understand the process of galaxy formation
and evolution, led to an increasing number of deeper redshift surveys.
Redshift surveys are now an ``industry" with its own standards.
Reduction of an ever growing number of data is based on software packages
specially developed to this aim.
The redshift $z= \Delta \lambda / \lambda$,
or, less rigorously, the ``recession velocity" $V = cz$, is
commonly determined using the wavelength shift of either absorption or 
emission lines appearing in the optical spectrum of a galaxy. 
Following the paper by Tonry \& Davis (1979), most redshifts based on
absorption lines are now obtained by cross--correlating galaxy spectra
with one or more (or an average of) ``template'' spectra, while
redshifts based on emission lines are measured by fitting the individual
emission lines. Moreover, emission and absorption lines are produced in 
different environments. In normal galaxies, the former (such as
the [OII]$\lambda 3727$ line) are generated in HII regions associated with 
recent star--formation, while the latter (such as the calcium
Ca II $K$ and $H$) are produced in stellar atmospheres and are related
to the bulk of the star population. As a consequence, 
emission and absorption redshifts are not required to be exactly the same. 

Despite the growing number of galaxy redshifts in the literature,
most catalogues quote only the ``best'' estimate of the velocity of a galaxy,
and take for granted the implicit and widespread
assumption that, while for a given galaxy
the absorption velocity $V_{abs}$ and the emission velocity $V_{emi}$
may differ, the average difference should be consistent with zero.

In the analysis of the ESO Slice Project (ESP; Vettolani et al. 1997;
Zucca et al. 1997; Vettolani et al. 1998), we have devoted a particular
effort to check the quality of our data, and in particular the precision
of our absorption and emission redshift measurements which, as we have
soon realized, present a puzzling discrepancy. 
Looking at the past and
recent literature, we have also realized that this problem was not new,
but was never discussed in a satisfactory way.
We have therefore decided to study the effect in more detail,
and we describe in this paper the results of our analysis
and the possible explanations.

In section 2 we discuss the evidence of discrepancies in
\dv~ found in the past and in other surveys, and in section 3 we 
present the discrepancy detected in the ESP data.
In section 4  we describe
the tests we have performed on the ESP data, exploring 
instrumental and other effects which could in principle
affect our results; in section 5 we analyse in detail the biases on
velocity measurements due to the choice of the template spectra;
in section 6 we discuss if such a discrepancy can be partly due to a real,
physical effect; our conclusions are in section 7.

\section{Systematic differences between absorption and emission 
line redshifts in previous surveys}

Systematic differences in redshift measurements
have been detected and discussed in the past. For example,
Roberts (1972) found a systematic difference between the HI 
and the optical redshifts of galaxies 
in the velocity range between 1200 and 2400 km/s,
which he attributed to the blending of galaxian and sky Ca II $H$ and $K$ 
absorption lines.
A small effect in the same range of velocities was also found by 
Sandage (1978), with $<V_{HI} - V_{emi}> = -33 \pm 22$~km/s and  
$<V_{HI} - V_{abs}> = -103 \pm 37$~km/s, i.e. a positive difference between 
absorption and emission velocities corresponding to about +70 km/s.
Sandage applied then a zero--point correction of +30 km/s to the
redshifts, for consistency with HI velocities,
but Tonry \& Davis (1979) found that the redshifts of the galaxies they had 
in common with Sandage were consistent with Sandage redshifts only if his
correction was not applied.

These puzzling results were not isolated. Corwin \& Emerson (1982)
analysed the spectra of 71 galaxies, and for 24 galaxies with both
absorption and emission velocities they found 
$<\Delta V > = <V_{abs}-V_{emi}> = +64 \pm 16$~km/s (hereafter
$\Delta V$ will always indicate the difference 
$V_{abs} - V_{emi}$).
Lewis (1983) found a zero--point error of about 30 km/s in the
data of Shectman, Stefanik \& Latham (1983), for which 
\dv~ was systematically negative. Interestingly enough,
when regarding the cases with the largest residuals,
he found that the $V_{HI}$ velocity was nearer to the $V_{abs}$ and
also that \dv~ was systematically positive. He concluded that these
cases ``[...] are most probably explained as large gas outflows from
the nucleus''. Similarly, 
Mirabel \& Sanders (1988) measured $<V_{HI} - V_{opt}> \sim 87$ km/s,
where $V_{opt}$ refers to emission line velocities,
for a sample of ultra--luminous dusty IRAS galaxies; they 
concluded that ``The discrepancy could be due to optical
line--emitting gas moving radially, probably outward,
in the central regions of luminous infrared galaxies.
If such outwardly moving emitting--line gas is mixed with dust, the
attenuation of emission from the far side leads to an observed optical
redshift below systemic.''
Similar results have been found for the [OIII] $\lambda$5007 line in
the Narrow Line Region of AGNs (see Wilson \& Heckman 1985 and section 4).

The above discussion shows that 
a) non--negligible systematic zero--point 
   differences are a common problem in redshift surveys;
b) sometimes, a systematic difference may be due to physical reasons, as
   in the case of ultraluminous infrared galaxies; 
c) with small number of galaxies, it is difficult to determine
   the amplitude and the reasons of the difference.

A better analysis of this problem can be done with larger samples.
For the main surveys where both absorption and emission line redshifts
are available, we show in table 1 the acronym of the survey (column 1),
the number $N_g$ of galaxies in the
sample (column 2), the limiting apparent magnitude (column 3) and
the mean $<V_{abs} - V_{emi}>$ with its error (column 4).
At relatively bright magnitudes ($b_J \le 17$), we have 3 main redshift surveys
where both $V_{abs}$ and $V_{emi}$ are available for more than $100$ galaxies: 
the Anglo--Australian Redshift Survey (AARS; Peterson et al. 1986), 
the South African Astronomical Observatory 
Redshift Survey (SAAO; Menzies et al. 1989), and the Stromlo--APM 
Redshift Survey (Loveday et al. 1996).
As apparent from table 1, the first two surveys do not show any 
systematic difference between $V_{abs}$ and $V_{emi}$.
Loveday et al. (1996) find $< \Delta V > \sim -19$~km/s for the
Stromlo--APM redshift survey, and conclude that this value ``is negligible
compared with the rms difference of 124 km/s". Indeed 
in the literature the average value of $\Delta V$ 
is often compared only with the rms of the $\Delta V$ distribution.
However, as Loveday et al. (1996)
have 825 galaxies with reliable absorption and emission velocities, 
this implies a standard error on the mean of about 4 km/s, i.e.
the value of $-19$ km/s formally differs from zero
at more than $4 \sigma$ level. Such a
systematic effect is obviously negligible, but in other cases it is
not, and it is important to keep distinct the rms of the $\Delta V$ 
distribution from the standard error of the mean.

There is indeed another source of redshifts, which gives a somewhat different
result: it is the redshift catalogue for
a magnitude limited sample ($b_J \le 16.5$)
obtained with the FLAIR multi--object spectrograph 
(Parker \& Watson 1990). For a sample of 80 galaxies,
the measures of redshift made on the red part of the spectrum give 
$ <\Delta V> = +52$ km/s, with rms $21$ km/s: this means that
the discrepancy is significant.
Among various tests, Parker \& Watson show that velocity
measures of the night sky emission lines are systematically shifted of 
about $-15$ km/s. On the other hand, absorption velocities
are measured by cross--correlating with only one template and,
as they notice, an error of $\sim 50$ km/s on the published velocity of the   
template could explain the difference.
For these reasons, it is not possible to prove the existence of a
real discrepancy in the FLAIR data.

The situation is different when looking at 
results obtained at fainter magnitudes.
For the Durham/Anglo--Australian Telescope faint galaxy redshift survey
(Broadhurst et al. 1988, hereafter BES), with an apparent
magnitude limit $b_J \le 21.5$, 97 galaxies have both absorption
and emission velocity.
While Broadhurst et al. do not discuss the problem,
from their published velocities and their quoted redshift
precision of $\sim 100$ km/s, we find a systematic difference
\dv $\sim +129 \pm 10$ km/s.

\begin{table}
\caption[]{Mean $\Delta{V} =$ \dv~ for various recent redshift surveys}
\label{tab:surveys}
\begin{flushleft}
\begin{tabular}{lrrrr}
\hline
Survey & N$_g$ & $b_J$ & $<V_{abs} - V_{emi}>$ (km/s) \\
\hline 
SAAO           & 142 & $\sim15$ & ~~-7 $\pm$ 7 \\
AARS           & 165 & $\sim17$ & ~~-2 $\pm 4$ \\
STROMLO--APM   & 825 & $17.15$  & ~-19 $\pm$ 4 \\
ESP            & 742 & $19.4$   & ~+94 $\pm$ 6 \\
BES            & ~97 & $21.5$   & +129 $\pm$ 10\\
\hline
\end{tabular}
\end{flushleft}
\end{table}

\section{\dv~ discrepancy in the ESP}

While the results of most previous surveys are based on a relatively small
number of galaxies, we have a catalog of $742$ galaxies
for which both emission and absorption line velocities
([OII]$\lambda$3727, 
$H_\beta$, [OIII]$\lambda$4959 \& $\lambda$5007) have been measured. 
This data set is a subsample of the recently completed ESO Slice Project
(for more details see Vettolani et al. 1997, 1998), a
statistically complete redshift survey of 3342 galaxies to a depth of
${b_J}=19.4$, selected from the Edinburg/Durham Southern
Galaxy Catalogue (Heydon--Dumbleton et al. 1988, 1989; Collins et al. 1989).
Observations were carried out at the 3.6m ESO telescope at La Silla,
with the multi--fiber spectrographs OPTOPUS
(Lund \& Surdej 1986) and MEFOS (Felenbok et al. 1997).
Exposure times were fixed, with 2 half an hour exposures for each field.
The spectral coverage of the survey ranges from
$3730$ \AA~ to $6050$ \AA, sampled at $\simeq 4.5$ \AA/pixel (corresponding
to $\sim 270$ km/s at 5000 \AA). 
Redshifts were determined using the {\em IRAF}\footnote{{\em IRAF} is 
distributed by the National Optical Astronomy Obser\-vatories, which is 
operated by AURA Inc. for the NSF.} 
external package {\em rvsao}, developed at the
Smithsonian Astronomical Observatory.
The absorption line redshifts were measured with the task {\em xcsao},
based on a cross--correlation technique, comparing the observed
galaxy spectra with 8 stellar template spectra, observed by us with the same
instrumental set up,  
and selecting the velocity given by the template with the smallest
error, while the emission lines were directly measured with the task 
{\em emsao}. The median internal velocity error for our data is $\sim 60$~km/s.
The appropriate heliocentric correction was applied to all velocities.

In the cross--correlation technique, the quality of a spectrum can
be judged by its $R$ parameter, defined as:
\begin{equation}
R = \frac{h}{\sqrt{2}\sigma_a}
\end{equation}
where $h$ is the height of the true cross--correlation peak and 
$\sqrt{2}\sigma_a$ is the height of an average, noise peak. 
The mean error on the measured shift of the spectrum, 
binned on a logarithmic scale, is $\propto (1 + R)^{-1}$;
therefore a larger value of $R$ generally corresponds to a lower error
(see Tonry \& Davis 1979 for a detailed discussion).

The task {\em emsao} finds emission lines, determines the peak wavelength  
of each identified line through gaussian fitting and computes its
redshift. If more than one emission line redshifts are measured,
{\em emsao} combines them into a single radial velocity. In our process of
data reduction, each galaxy spectrum was carefully checked by eye, 
in order to avoid spurious identifications.

We carefully examined the galaxies with the largest \dv~ in the ESP survey 
before building the final catalogue. Apart from a few cases where
the difference was obviously spurious, due to an error in writing one velocity,
or in the identification of an emission line, most of the remaining
large $\Delta V$ were confirmed, and in these cases 
a genuine physical explanation is probably required
(see section 4).

We find for 742 galaxies
$<\Delta V> = 93.7 \pm 6.1 $~km/s, with a standard deviation 
$\sigma = 166$ km/s, while the corresponding weighted estimates are
$<\Delta V> = 90.8 \pm  5.0$~km/s and $\sigma = 138.5$~km/s,
with no statistically significant skewness.
We point out that that the formal error values of the IRAF tasks 
{\em xcsao} (for absorption
line velocities) and {\em emsao} (for emission line velocities)
were multiplied respectively by the factors 1.53 and 2.10, in order to obtain 
estimates of the true errors (these corrections are based on the
analysis of galaxies observed more than once, see Vettolani et al. 1998).
In this paper the weighted estimates and the associated
errors are always computed applying these correction factors.

The observed width of the $\Delta V$ distribution results in principle
from the convolution of the intrinsic width of the distribution
($\sigma_{int}$) with the measurement error distribution. Applying a
maximum likelihood technique (see Maccacaro et al. 1988), which takes
into account the error associated with each measurement, we find
$\sigma_{int} \sim 40$ km/s, with a three sigma allowed range from
0 to 68 km/s. 
Therefore, the effect responsible for
the positive \dv~ cannot have a very large intrinsic dispersion.

\begin{figure}
\resizebox{\hsize}{!}{\includegraphics{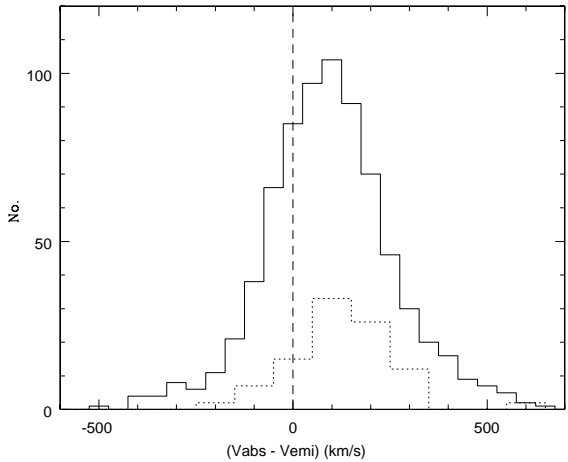}}
\label{fig:1}
\caption[]{Histogram of the velocity difference 
$\Delta{V} = V_{abs} - V_{emi}$; solid line: ESP; dotted line: BES.}
\label{fig:1}
\end{figure}

As shown in table 1, in the ESP and BES surveys the 
absorption line velocities are systematically 
higher than emission line velocities, in contrast to the shallower
surveys, where no discrepancy is detected.
In figure 1 we show the velocity histograms for the ESP and the BES,
where the $\Delta V$ asymmetry with respect to 0 (dashed line) is clear
(notice that the adopted bin for the BES is twice as large as for
the ESP data, since the number of objects in the BES survey is smaller). 

In order to investigate the possible systematic effects
which could give rise to the observed \dv~ in the ESP, we have carried out 
a large number of tests on our data.
Concerning the quality of the spectra, it is remarkable that
the velocity shift is confirmed --and even
larger-- for redshifts which have a high $R$ parameter, 
rising from $90.3 \pm 5.1$ km/s for $R \ge 2$ (694 galaxies), to 
$108.5 \pm 9.5$ km/s for $R \ge 7$ (100 galaxies).
Restricting the sample to the
54 galaxies showing at least 4 emission lines
and with $R \ge 3$ gives $< V_{abs}-V_{emi} > = +90.1 \pm 15.1$~km/s.

In table \ref{tab:app} we show in column 1 the limits defining
the subsample (respectively minimum and maximum apparent magnitude,
absolute magnitude, and redshift), in column 2 the number of galaxies
$N_g$, in column 3 the weighted average $< \Delta V >$ with its error, 
and in column 5 the rms.
From table \ref{tab:app} it appears that
the amplitude of the effect is smaller at brighter
apparent magnitudes ($b_J \le 18$), at lower redshifts ($z \le
0.08$), and at fainter absolute magnitudes ($M \le -19$).
The uncertainties are quite large, and it is not
possible to establish a significant trend, but the
data suggest an increase of the effect with distance and 
absolute magnitude.

As the effect is significant in all magnitude and redshift
bins, we can exclude that it is due to the 
blending with some specific sky lines.

\begin{table*}[ht]
\caption[]{Velocity differences as a function of apparent magnitude, redshift, 
and absolute magnitude for the ESP survey (weighted estimates).}
\label{tab:app}
\begin{flushleft}
\begin{tabular}{rrrr}
\hline
 $m$ & N$_g$  & $\Delta{V}$ & rms \\
\hline
 $15 \le m < 18$   &   142 & $71.6 \pm 9.6$ & 115.2 \\
 $18 \le m < 19$   &   350 & $98.0 \pm 7.3$ & 132.8 \\
 $19 \le m \le 19.4$ &   249 & $99.1 \pm 9.7$ & 166.2 \\
\end{tabular}
\begin{tabular}{rrrr}
\hline
 $z$ & N$_g$ & $\Delta{V}$ & rms \\
\hline
$0.00 \le z \le 0.08$ & 178 & ~$72.1 \pm 9.1$  & 107.2 \\
$0.08 \le z \le 0.12$ & 239 & ~$99.7 \pm 8.6$  & 133.5 \\
$0.12 \le z \le 0.16$ & 154 & $100.7 \pm 11.7$ & 154.3 \\
$0.16 \le z \le 0.24$ & 158 & $100.2 \pm 12.4$ & 176.5 \\
 \end{tabular}

\begin{tabular}{rrrr}
\hline
 $M$ & N$_g$ & $\Delta{V}$ & rms \\
\hline
$-18 \le M \le -16$ & ~67 & ~$84.9 \pm 16.4$ & 110.8 \\
$-19 \le M \le -18$ & 164 & ~$81.6 \pm 10.3$ & 121.4 \\
$-20 \le M \le -19$ & 278 & ~$93.4 \pm 8.1$ & 146.1 \\
$-21 \le M \le -20$ & 216 & ~$99.1 \pm 9.5$ & 149.1  \\
\hline
 \end{tabular}
 \end{flushleft}
 \end{table*}

\section{Possible instrumental errors}

\subsection{Velocities zero--point}

In the cross--correlation method, the most important point of concern
is of course the zero--point of the templates. Therefore we measured
the velocities of three SAO radial velocity standard stars;
averaging the values obtained from our 8 ESP templates, we
find  that the velocity estimates of the 3 stars agree with the 
literature values within a few kilometres per second.
Moreover, a comparison with 7 galaxies for which $HI$ velocities are available
shows that the mean zero--point of the 8 stellar templates is lower
than the mean HI velocities by $-17 \pm 10$~km/s.
In table \ref{tab:tempHI} we give the velocity difference $V_{ESP} -
V_{HI}$, where $V_{ESP}$ is the velocity obtained from the
cross--correlation and $V_{HI}$ is the literature velocity, for each template.

\begin{table*}
 \caption[]{$V_{ESP}$ - $V_{HI}$ obtained with our 8 templates on 7 galaxies
with $HI$ redshifts.} 
  \label{tab:tempHI}
 \begin{flushleft}
  \begin{tabular}{rrrrrrrrr}
\hline
Gal. no. & 1 & 2 & 3 & 4 & 5 & 6 & 7 & $<$V$>$\\ 
\hline
ESP template 1 & ~21 & ~-9 & ~~7 & ~~0 & ~39 & -29 & -52 & $ -3 \pm 11 $ \\
ESP template 2 & ~57 & -37 & ~40 & ~~1 & ~-4 & -29 & -46 & $ -3 \pm 15 $ \\
ESP template 3 & ~~6 & -33 & ~-2 & -20 & ~~3 & -31 & -55 & $ -20 \pm 8 $ \\
ESP template 4 & ~14 & -23 & ~-8 & -20 & ~26 & -24 & -51 & $ -12 \pm 10$ \\
ESP template 5 & ~21 & ~~2 & ~-1 & -19 & ~32 & -37 & -67 & $ -10 \pm 13$ \\
ESP template 6 & ~38 & -48 & ~24 & -17 & -16 & -35 & -52 & $ -15 \pm 13$ \\
ESP template 7 & ~15 & -50 & -27 & -60 & -19 & -34 & -62 & $ -34 \pm 10$ \\
ESP template 8 & ~18 & -17 & ~~1 & -21 & ~~5 & -36 & -60 & $ -17 \pm 10$ \\
\hline
  \end{tabular}
 \end{flushleft}
\end{table*}

Table \ref{tab:tempHI} shows that our 8 templates give consistent
results. The template which seems to
give the largest underestimate of the velocity (relatively to the $HI$
velocities), is no.7
 
The 9 galaxies with measured velocity we have in common with
the Stromlo--APM redshift survey (Loveday et al. 1996)
give a mean difference $V_{ESP} - V_{APM} = -7.9$ km/s.  
This means that we expect a negligible zero--point error, with possibly 
a small {\em underestimate} of the true absorption velocity: if we should
apply such a correction, the systematic difference between $V_{abs}$ and
$V_{emi}$ we find would be even larger.

\subsection{Wavelength calibration}

After excluding a zero--point error, the next candidate for an explanation
is a calibration problem. For example, we can suppose that 
in our wavelength calibration there is a systematic error between
the blue and the red parts of the spectrum, and that the velocity discrepancy
is due to the underestimate of the [OII]$\lambda 3727$ line redshift.
   
\begin{figure}
\resizebox{\hsize}{!}{\includegraphics{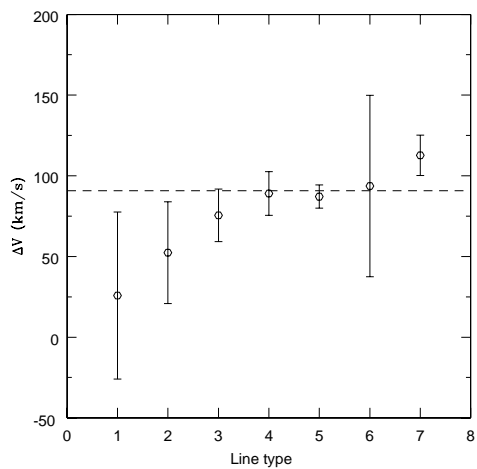}}
\caption[]{\dv~ (with $1 \sigma$ errors) as a function of the lines
present in the spectrum;
1 = $H_\beta$, 2 = $H_{\beta}$ + [OIII]a \& b, 
3 =  $H_{\beta}$ + [OII], 4 = $H_{\beta}$ + [OII] + [OIII]a\&b,
5 = [OII], 6 = [OIII]a\&b, 7 = [OII] + [OIII]a \& b. The dashed line shows
the average value found for the whole sample.}
\label{fig:2}
\end{figure}

Table \ref{tab:lines} shows how the error weighted estimate of
$<\Delta V>$ depends on the
emission lines detected in the spectrum (see also figure \ref{fig:2});
we have listed all the possible cases except when only one of the
two [OIII] lines is present.
 
From table \ref{tab:lines} we see that all emission lines give positive
~\dv, and that the values are all consistent, except for the $H_{\beta}$
and the [OII]$\lambda$3727 + 
[OIII]$\lambda$4959 \& $\lambda$5007 lines,
which give respectively the smallest and the largest velocity
difference.

The most important implications of table \ref{tab:lines} are that 
a) the effect does not depend only on [OII], thus excluding
a mismatch due to the particular shape of the line (as [OII] is
in fact a doublet which is not resolved in our spectra);
b) we have a very good calibration both in the blue and in the red part of the 
spectrum.

\begin{table}[ht]
\caption[]{Error weighted \dv~  
as a function of the emission lines detected in the spectrum}
\label{tab:lines}
\begin{flushleft}
\begin{tabular}{lrrr}
\hline
 Lines                            & N$_g$ & $\Delta{V}$ & error \\
\hline
 $H_{\beta}$                      & ~~8 & ~26 & 52 \\ 
 $H_{\beta}$ + [OIII]a \& b       & ~22 & ~52 & 32 \\
 $H_{\beta}$ + [OII]              & ~68 & ~76 & 16 \\
 $H_{\beta}$ + [OII] + [OIII]a\&b & ~88 & ~89 & 14 \\
\ [OII]                           & 430 & ~87 & ~7 \\
\ [OIII]a\&b                      & ~~7 & ~94 & 56 \\
\ [OII] + [OIII]a\&b              & ~64 & 113 & 13 \\
\hline
  All                             & 742 & ~91 & ~5 \\
\hline
\end{tabular}
\end{flushleft}
\end{table}

For the 8 galaxies where only the $H_\beta$ emission
line has been detected, the mean emission velocity is formally consistent 
with the absorption velocity, but is also consistent with the total sample
average at the $2 \sigma$ level. We stress that
the $H_\beta$ emission line shape is not easy to fit.
We point out also that the absorption $H_\beta$ rest wavelength usually assumed
for galaxies (and for example used in $IRAF$) is at 4863.9 \AA~ 
i.e., shifted by 2.6 \AA~ with respect to the laboratory wavelength. 
This value was first given by
Sandage (1975; the Sandage values are also reported by Loveday et al. 1996)
and implicitly attributed to the blending with
other lines at low resolution. We do not know of more recent tests
about this shift, which should depend on the instrumentation and
the galaxy type. 
It is worth noticing that, in order to find the ``effective''
wavelengths of blended lines in galaxies, Sandage explicitely assumed that
the emission lines indicate the true velocity of the system; any
systematic velocity of the emission lines relative to the galaxy velocity
would be reflected in the values of the absorption wavelengths.   

With the absorption $H_\beta$ line at a slightly larger wavelength than the 
emission $H_\beta$, the velocity of the latter might be
underestimated, which would justify the small difference with the other 
emission lines.
On the other hand, supposing that in our case the effective rest 
wavelength of the $H_\beta$ absorption line is smaller
than the value given by Sandage (1975),
--i.e. nearer to the laboratory wavelength-- 
then assuming the Sandage value
we would underestimate the absorption $H_\beta$ line
redshift and, consequently, the amplitude of $\Delta V$.

Table \ref{tab:lines} tells us that if such effects are present in our sample,
they are not large.

\subsection{Other possible errors}

We have taken into account many other possibilities.

It is well known that the [OII]$\lambda3727$ line is in fact 
a doublet, unresolved at our
resolution. We have checked that this cannot significantly affect
our fit and, as we have shown, the discrepancy is present also
with the other emission lines.

Before cross--correlating a spectrum, the emission lines
of the galaxy, if present, were removed. 
Within {\em xcsao} it is also possible to perform this task
automatically, but we decided to remove emission lines manually,
after a comparison of the results obtained with the two methods
(varying also the parameters in the automatic procedure), as we found that 
the typical difference in the velocity is $\sim 20$ km/s, with no
evidence of a systematic overestimate with our method.

In the cross--correlation method the continuum is fitted
and subtracted from the spectrum. Therefore we have changed the fitting 
functions and orders, without any significant variation in the results.
These tests cannot exclude a systematic difference due to the way
{\em xcsao} normalizes the continuum; but the fact that the effect
is present at very different redshifts suggests that this
is not the case.

The fact that the velocity difference is significant in 
all magnitude and redshift
bins (see table \ref{tab:app}) excludes also that it may be due to the 
blending with some sky lines (as in the case of the Roberts effect).

Before concluding that the velocity discrepancy between
absorption and emission lines is real, we have still to focus
on an important aspect of the cross--correlation technique.
Such technique relies on information external to the data:
the template spectra. In the cross--correlation method 
we assume that the best redshift
measurement is given by the best--fitting template. In the next
section we will carefully check the validity of this assumption. 

\section{Dependence of the redshift measurements on the template spectra}

In addition to the ESP, another recent
redshift survey presents a systematic discrepancy between absorption
and emission line redshifts: it is the Las Campanas Redshift Survey
(LCRS; Shectman et al. 1996).
When discussing the data reduction for the LCRS,
Shectman et al. (1996) note that a systematic bias ``creeps into 
cross--correlation velocities for the emission line galaxies".
They ascribe this bias to systematic differences
in the absorption--line spectra of standard templates
(which are usually late--type stars) and emission line galaxies, in particular
to a blend between the $H_\epsilon \lambda 3970$ and Ca II $H$ lines.
This blending was already tabulated
in the classical work by Humason, Mayall \& Sandage (1956),
and described by Sandage (1978). Humason et al. (1956) 
list 3 possible effective wavelengths for the
blend between $H_\epsilon$ and the Ca~II $H$ line, depending on their
relative intensities: 3968.54\AA, 3969.01\AA~ and 3969.23\AA, 
with the Ca~II $H$ rest wavelength
line at 3968.38\AA. Sandage (1978) discusses the dependence
of the blending on the galaxy type, concluding that it can be of the
order of $\sim 100$ km/s.
 
Indeed using a template with strong Balmer lines, Shectman et al.
(1996)  reduce the systematic effect to about 15 km/s.
Therefore they use this template for galaxies with emission lines and 
two standard templates for all the other galaxies.
However, we think one should be careful in taking this
possibility as a definitive explanation, only on the basis of the 
{\em a posteriori} agreement between $V_{abs}$ and $V_{emi}$.
It would be surprising that the results of the
cross--correlation should depend so critically on one line
(even if the $H$ line is surely important), and other lines should 
conspire to give this difference (as suggested by Shectman et al. 1996).

\begin{figure}
\resizebox{\hsize}{!}{\includegraphics{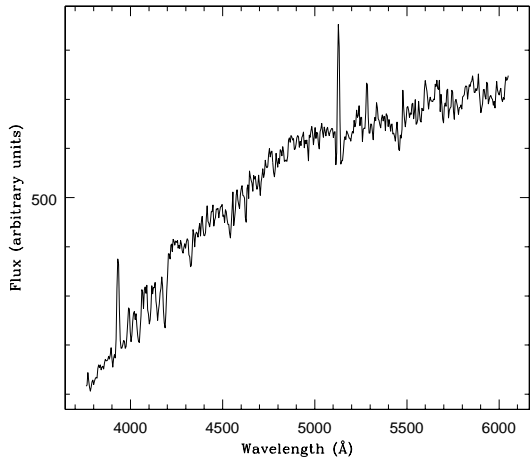}}
\caption[]{Spectrum of a galaxy with $V_{abs} - V_{emi} \sim 30$ km/s.}
\label{fig:3}
\end{figure}

We performed some tests cross--correlating the 3 LCRS templates 
(kindly provided by H.Lin) together with
our 8 star templates to our spectra.
We confirm that the LCRS template with strong Balmer lines gives
systematically lower velocities than the other ones.
However, the templates which give the best--fit to galaxy spectra
are systematically the standard ones, while the template with strong
Balmer lines usually gives a lower $R$, 
even in cases where there are strong emission lines.
Moreover, such template should give lower absorption redshifts
only for galaxies with a significant discrepancy
between $V_{abs}$ and $V_{emi}$, while
it gives lower redshifts also for other galaxies.
For example, in the case of the spectrum shown in figure \ref{fig:3}, 
% [93oct032 aperture 10]
the standard Shectman templates perform better, and are consistent
with our 8 templates, while the non--standard Shectman template performs
more poorly, and gives a velocity about 100 km/s lower:
but in this spectrum, the emission line velocity is
consistent with the absorption one (only 30 km/s lower), so applying
the special template to this emission line spectrum would in fact
underestimate the redshift!
This illustrates the main risk of using a different template for emission
line galaxies: it may solve the \dv~ problem in a statistical
sense, but it might not work with the individual galaxies. 

However, this does not exclude the possibility that a systematic effect 
can be introduced by the choice of a template.
In order to clarify the issue, we have compared
the velocities given by {\em all} our 8 templates 
for each galaxy in a subset of 8 ESP fields.
The results are shown in table \ref{tab:templates}, where we report
in the first column the ESP field number, and in columns
from 2 to 7 the mean difference $<V_1 - V_J>$, $J=2,8$, between
the velocities measured with the ESP template no.1 
(arbitrarily chosen as a reference) and with each one of the other 7 
ESP templates.

\begin{table*}[hb]
\caption[]{Average cross--correlation velocity differences $\Delta V_{cc}$ 
measured in a subsample of ESP galaxies with our 8 templates. 
The velocity obtained with the ESP template no.1, $V_1$, is conventionally 
taken as a reference.}
\label{tab:templates}
\begin{flushleft}
\begin{tabular}{lrrrrrrr}
\hline
ESP Field & $<V_1-V_2>$ &  $<V_1-V_3>$ &  $<V_1-V_4>$ &  $<V_1-V_5>$ &  $<V_1-V_6>$ &  
$<V_1-V_7>$ &  $<V_1-V_8>$ \\ 
\hline
                 &       &       &        &       &         &       &     \\
 Field 104             &       &       &        &       &         &       &     \\
 $N_g$           & 24    &  30   &   29   &   28  &  27     &   29  & 28   \\
 $\Delta V_{cc}$ & 11.6  & -17.9 &  -52.6 &   9.3 &  15.4   & -45.8 & 8.4 \\ 
 Field 106             &       &       &        &       &         &       &     \\ 
 $N_g$           &  9    &   16  &   19   &   16  &     13  &   17  & 18 \\
 $\Delta V_{cc}$ & 41.2  &  -7.6 &  -55.6 & -19.2 &    64.5 & -33.7 & 4.8 \\
 Field 107             &       &       &        &       &         &       &     \\
 $N_g$           &    22 &  29   &   29   &   28  &      24 &   30  &  30 \\
 $\Delta V_{cc}$ &  38.9 &  1.4  &  -29.2 &  15.3 &    42.7 & -15.2 &  20.2 \\
 Field 108             &       &       &        &       &         &       &       \\
 $N_g$           &    22 &   28  &     29 &   30  &     22  &   29  &  29 \\
 $\Delta V_{cc}$ &   8.6 & -15.0 &  -38.8 &  13.0 &    12.7 & -20.9 &   8.4 \\
 Field 109             &       &       &        &       &         &       &       \\
 $N_g$           &    12 &   22  &     22 &  19   &    14   &   22  &  22 \\
 $\Delta V_{cc}$ &  37.4 & -0.7  &  -32.1 &  29.9 &    56.5 & -14.8 &  13.9 \\
 Field 121             &       &       &        &       &        &        &      \\
 $N_g$           &    20 &   31  &     31 &  31   &     22 &    31  &  32 \\
 $\Delta V_{cc}$ &  -8.5 & -8.1  &  -35.7 & 10.4  &   12.9 &  -13.4 &  4.4 \\
 Field 145             &       &       &        &       &        &       &       \\
 $N_g$           &    12 &   16  &   16   & 16    &   14  &   15  &   16 \\
 $\Delta V_{cc}$ &  49.4 & -8.3  &  -57.6 & 5.7   &  51.2 & -35.2 &  -2.2 \\
 Field 164             &       &       &        &       &        &       &       \\
 $N_g$           &    20 &    23 &    23  &  23   &    22  &    23 &   22 \\
 $\Delta V_{cc}$ &  14.9 &  5.7  &  -37.4 & -9.5  &   34.4 &  -7.9 &  17.4 \\
\hline        
                 &       &       &        &       &         &       &     \\
Total $N_g$      &  141  & 195   & 198    &  191  & 158     & 196   & 197  \\
$<\Delta V_{cc}>$ &  $20 \pm 7$   & $-7 \pm 3$     & $-41 \pm 4$  & $8 \pm 5$
                 & $ 32 \pm 7$   & $-23 \pm 5$    &  $10 \pm 3$ \\
                 &       &       &        &       &         &       &     \\
\hline
\end{tabular}
\end{flushleft}
\end{table*}

Table \ref{tab:templates} shows in a clear way that systematic differences
do exist: for example, templates no.4 and no.7 give on the mean larger 
velocities.
The systematic effect between template no.4, which gives
the largest velocity, and template no.6, which gives the lowest velocity,
is of about 70 km/s. Notice that this {\em cannot be due} to a
zero--point error. This becomes very clear when comparing
table \ref{tab:templates} with table \ref{tab:tempHI}, which shows
the zero--point shifts of the ESP templates as estimated from a comparison 
with the $HI$ velocities of 7 galaxies. From table \ref{tab:tempHI}
we see that the ESP template no.7 
gives the lowest estimate of the HI velocities, and we would
expect $<V_1 -V_7> \sim +30 \pm 15$ km/s instead of
$<V_1 -V_7> \sim -23 \pm 5$ km/s;
for template number 4, from our comparison with $HI$
velocities we would have expected $<V_1 - V_4> \sim +9 \pm 15$ km/s,
instead of $<V_1 - V_4> \sim -41 \pm 4$ km/s!

We conclude that different templates give systematically different 
velocities, probably because each template fits in a different way 
the observed galaxy spectra. In the light of our results,
the velocity difference between the 3 templates by Shechtman et al. (1996)
appears to be a particular case.

Can the results shown in table \ref{tab:templates} 
explain our $<\Delta V>$ problem?
This would be the case if the templates no.4 and no.7 gave a
systematically smaller error (as in the ESP we have attributed to each
galaxy the velocity of the best--fitting template).
We have indeed found that these two templates were used for 47\% of
the galaxies in the subset of fields shown in table \ref{tab:templates}.   
However, replacing velocities obtained with the two 
templates giving systematically larger velocities,
with the best one obtained by any one of the other templates, 
the resulting \dv~ is $\sim +65$ km/s. The discrepancy is reduced,
but still present; and should we after all rely on templates which
perform worse but which give a ``better'' result?

\begin{figure}[h]
\resizebox{\hsize}{!}{\includegraphics{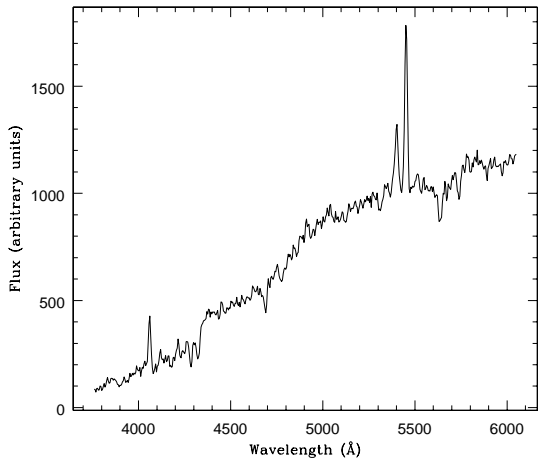}}
\caption[]{Spectrum of a galaxy with $V_{abs}-V_{emi} \sim 200$ km/s.}
\label{fig:4}
\end{figure}

That things are not so simple is shown by another example.
We have chosen a spectrum with high signal--to--noise ratio
(figure \ref{fig:4}), showing a difference $V_{abs}-V_{emi}$ of 
$\sim 200$ km/s, and we have looked at the values given by our
templates and by the Shectman templates (table \ref{tab:templates2}).
\begin{table}[ht]
\caption[]{Results of the cross--correlation for the spectrum of figure 4.}
\label{tab:templates2}
\begin{flushleft}
\begin{tabular}{lrrr}
\hline
Template         & Velocity & Error & R \\
\hline
ESP template 2    & 26822 & 32 & 11.8 \\
ESP template 8    & 26858 & 33 & 11.7 \\
ESP template 6    & 26850 & 35 & 11.1 \\
LCRS std templ.   & 26831 & 35 & 10.5 \\
ESP template 1    & 26838 & 37 & 10.0 \\
ESP template 4    & 26837 & 44 & ~8.9 \\
ESP template 7    & 26831 & 45 & ~8.6 \\
ESP template 3    & 26793 & 39 & ~8.2 \\
LCRS A template   & 26798 & 78 & ~4.1 \\
ESP template 5    & 20141 & 83 & ~2.0 \\
\hline
\end{tabular}
\end{flushleft}
\end{table}
It is clear that the template with strong Balmer lines (LCRS A template)
gives a lower value, but not sufficient to explain the difference;
moreover, it performs very poorly.
On the other hand, when we fit the main absorption or emission lines 
with a gaussian, we find the values shown in table \ref{tab:5}.
\begin{table}[h]
\caption[]{Results from gaussian fitting of the absorption and emission
lines in the spectrum of figure 4.}
\label{tab:5}
\begin{flushleft}
\begin{tabular}{lrr}
\hline
Line       & $\lambda$ & Velocity \\
\hline
\ Ca II $K$  & 4284.8~ &  26758 \\
\ Ca II $H$  & 4323.15 &  26791 \\
\ G          & 4688.31 &  26739 \\
\ Mgb        & 5636.76 &  26728 \\
\ FeCa       & 5738.75 &  26729 \\
\ [OII]      & 4059.6~ &  26727 \\ 
\ [OIII]4959 & 5399.73 &  26650 \\
\ [OIII]5007 & 5452.55 &  26688 \\
\hline
\end{tabular}
\end{flushleft}
\end{table}
In this case the cross--correlation has apparently overestimated
the redshift by $\sim 100$ km/s relatively to the gaussian fitting
measure. It can be noticed that neither 
template no.4 nor template no.7 are among the 3 best--fitting
templates, and template no.6 gives a relatively high velocity.
We still find a positive $\Delta V \sim 100$ km/s; however, 
if we could generalize this result, by concluding that the cross--correlation
overestimates the redshift by $100$ km/s, 
the velocity bias would obviously vanish. 
The problem is that for spectra with a lower S/N
ratio the direct measure of absorption lines is much more uncertain,
and when the velocity difference between absorption and emission lines
is around 100 km/s the Gaussian fitting of a few lines is not 
sufficiently accurate.

\begin{figure}
\resizebox{\hsize}{!}{\includegraphics{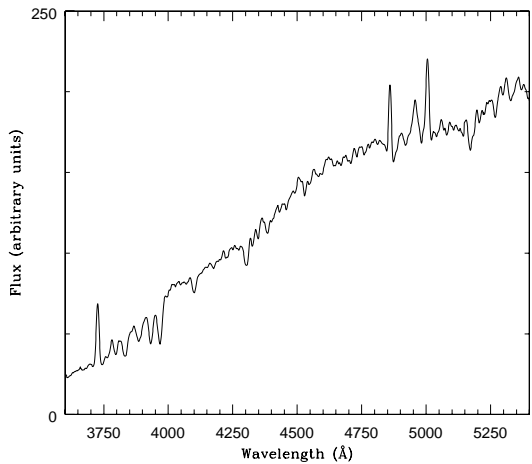}}
\caption[]{Coadded spectra of 16 galaxies, after having shifted to
 rest wavelengths according to their measured {\em absorption line} velocity.}
\label{fig:composite}
\end{figure}

We tried to avoid this problem by selecting a subset of 16 spectra with
a large $R$ parameter, most of them with 4 emission lines, and with a positive
$\Delta V$: this subset has an average $\Delta V = 124$ km/s.
We have shifted these spectra to the rest wavelength expected from
their measured {\em absorption line} velocity, as given by {\em
  xcsao}; then we have added these
spectra, to build a composite spectrum with higher S/N ratio. We show
this spectrum in figure \ref{fig:composite}.
Before addition, the individual spectra were resampled to 2048 pixels.
We have measured the wavelengths of the main lines through a gaussian fit.
We find that the emission lines are systematically
blueshifted: the [OII]3727 by 
$-80$ km/s, the $H_{\beta}$ by $-99$ km/s, and
the 2 [OIII] lines respectively by $\sim -147$ km/s and $-148$ km/s.
The [OIII] lines show also an asymmetric ``bump'' in their blue tail.

On the other hand, the wavelength of the Ca II $H$ line is at $3968.5$ \AA,
corresponding to zero km/s.
However, all the other absorption lines are also blueshifted.
For example, the Ca II $K$ line has a shift corresponding to 
-96 km/s. The $H_\delta$ line gives $-73$ km/s, the G--band $-70$ km/s,
the FeCa -39 km/s.
The absorption $H_{\beta}$ and Mg$b$ lines are also present,
but they are quite asymmetric, so that a gaussian fit cannot be a
reliable measure.

Computing the average velocity of the selected lines with their
errors, we find $<V_{abs}> = -56 \pm 17$ km/s including the Ca II
$H$ line, or $<V_{abs}> = -70 \pm 12$ km/s excluding it, and 
$<V_{emi}> = -119 \pm 17$ km/s.

Before drawing strong conclusions, we should emphasize that the
Gaussian fit is not very accurate, as it depends on the estimate by
eye of the continuum; moreover, the spectra have been chosen for their
strong absorption lines, which implies that they are not among those
spectra with the strongest emission lines, and might not be
representative of the total sample; finally, the effect will be
somewhat diluted as each galaxy has a different $\Delta V$.

We have also applied the cross--correlation method to the composite
spectrum. The formally lowest errors are obtained with the standard Shectman
template, giving $V = -20$ km/s ($R =10.6$), and our template no.8, 
giving $V = 14$ km/s ($R = 9.0$). The LCRS A template, with strong
Balmer lines, performs poorly ($R = 6.56$) but gives $V = -72$ km/s, a
measure in good agreement with the estimate derived from the gaussian
fit of the lines..

With all the uncertainties we have mentioned, it seems that the
cross--correlation overestimates the redshift,
apparently giving a large weight to the CaI $H$ line near the 4000\AA~
break, thus confirming, at least in part, the suggestion by Shectman
et al. (1996). 
It appears that in our case the effective rest wavelength
should be fixed to about $3869.5$ \AA
~(comparable to the third wavelength value for this line listed by Humason et
al. 1956), instead of the usual 3968.5 \AA.

We cannot conclude, however, that this effect can completely
explain the discrepancy with the emission line redshift.
Notice in fact that in the average spectrum the two [OIII] emission lines are
still discrepant by about 70 km/s relatively to the average of the
absorption line velocities (excluding the Ca II $H$ line).

Moreover, as the above examples have shown, for a given galaxy
we have no way to decide {\em a priori} which
is the template giving the best estimate, if
we leave the best--fit criterium.
For the above reasons, we decided to measure the 
redshifts of all spectra in the ESP with the technique of the 
best--fitting template, without forcing a given template to fit a given 
class of galaxies.

On the other hand, the choice of the template which gives velocities in
agreement with emission line velocity means we assume
that the system velocity is given --on the average-- 
by the emission line velocity.
In the absence of a definitive understanding of the bias,
we think it is important to check that emission lines
are on the average at rest, and that they cannot have
at least partially, a physical origin. We 
will briefly consider this possibility in the next section.

\section{Could it be a real effect?}

A relative shift of absorption and emission lines can
be a consequence of significant infall or outflow of gas in a galaxy. 
Cowie et al. (1995) in fact find evidence of infall of gas from their
observations of distant galaxies. 
Assuming that at least part of the velocity shift detected in the ESP 
is real, we have to find a relatively common process internal to  
galaxies which can give rise to ``peculiar motions'', with outflow   
 of the line emitting gas of $\sim 100$ km/s.

As briefly mentioned in the introduction,
a similar effect exists for the Narrow--Line
Region of AGNs, consisting in a systematic blueshift of
the [OIII]5007 line velocity relatively to the HI or absorption line velocity, 
and amounting to about $50$--$100$ km/s (Wilson \& Heckman 1985),
with the difference ranging between -250 and 250 km/s. It can be
due to the asymmetry of the line, generated by an outflow component
on the near side of the AGN (see Peterson 1997).

It is known that emission lines from normal galaxies tend to trace the
young stars formed in HII regions located on the spiral arms of late--type
galaxies, while the absorption lines dominate the stellar light from the
inner bulge component of galaxies (Vogel et al. 1988;
see also Knapen et al. 1992). However, there is no evidence that
the velocity bias we see is connected to the rotation of galaxies as there is 
no clear correlation between $V_{abs}-V_{emi}$ and galaxy inclination, as
we could determine from the axial ratios. 

While individual HII regions usually have thermal expansion velocities 
of $10-20$ km/s, which is insufficient to
account for the result, there is growing evidence of substantial 
amounts of diffuse ionized gas at ${T}\sim10^6$ K, shock heated by either 
supernovae or stellar winds, in some galaxies; this gas can have 
expansion velocities $\ge 100$ km/s after $10^5$ yrs (Spitzer 1990). 
Such gas has been conjectured to contribute as much as 
$50\%$ of the line emission in some star-forming galaxies and can
propagate to scale-heights ($\simeq5$ kpc) above the galaxy disk forming 
``galactic fountains'' of gas. As already noted,
velocity shifts of 100 km/s between optical emission line
galaxies and 21cm observations of the neutral gas component have been
found in ultra--luminous dusty IRAS galaxies (Mirabel \& Sanders 1988). 
In these systems
the effect is explained in terms of line-emitting gas moving radially
outward in the central regions of such galaxies. The presence of large
quantities of dust, mixed with the gas, provides the necessary
attenuation of the emission from the far side of the galaxy.

In the ESP survey we can only identify a handful of Seyferts and since
the survey is a complete optically selected sample, the
velocity bias should be a more common feature of the overall galaxy population 
than was previously thought and not simply confined to extragalactic objects 
with exotic properties.

This could be related to a steady increase in the amount
of turbulent gas within galaxies with look--back time. Radial motions of 
gas could have a significant role to play in galaxy evolution and may well 
provide a new probe of the early history of galaxies.
On the other hand, at a fainter limiting apparent magnitude
the fraction of intrinsically faint galaxies can increase due
to evolution, and one could alternatively imagine that a population of 
dwarf, star--forming galaxies is responsible for the effect.

The fact that the  $<V_{abs}-V_{emi}>$ may be larger at brighter
absolute magnitudes and redshifts
(see table \ref{tab:app} for the ESP and table \ref{tab:surveys} for
a comparison with other surveys) might be more consistent with
an evolution with redshift.

\section{Conclusions}

We have discussed in detail the errors associated to redshift determinations
in the ESO Slice Project (Vettolani et al. 1997). 
We have found a systematic difference between
absorption and emission line velocities
of $\sim +100$ km/s and
we have shown that the same effect is present in the
Durham/Anglo--Australian Telescope faint galaxy redshift survey 
(Broadhurst et al.1988).
In the case of the ESP, we have excluded problems of zero--point error
or calibration.
Such a discrepancy has not been detected in large shallower surveys.

Shectman et al. (1996) briefly discuss
a similar effect for the Las Campanas Redshift Survey, 
which they have corrected by using a different template
for emission line galaxies.

We have generalized the suggestion by Shectman et al. (1996), 
who identify as the main cause of the discrepancy
the systematic difference between the absorption line spectra of the
standard templates and the typical emission line galaxy, particularly
the blend between the Ca II $H$ and $H_\epsilon$
lines, an effect already discussed by Sandage (1978). 
We find in fact systematic effects even 
from template to template, apparently due to the way each template fits 
the galaxy spectra; this implies that the choice of the template 
significantly affects redshift measurements.

For the ESP data, we decided to use the best--fit
template (i.e. the one giving the smallest error), as using a different
template might introduce unknown biases in the redshift measurement.

In the lack of a definitive explanation, the common assumption
that the true galaxy redshift is given -- on average -- 
by the emission lines, is plausible, but not proven.
It should be verified that the bias cannot be due, at least partially,
to emission lines, and that the sample is not biased, for some reason,
towards galaxies with outflows.
One can for example speculate that other factors may contribute to the 
effect, such as sampling of different parts of the galaxies, the different 
mix of morphological types, and evolution with look--back time. 
A collection of high resolution data of a sample of galaxies 
and of different templates will be necessary to give a definitive solution.

We feel it is important to stress the existence of such an ``anomaly":
in view of  future, large surveys,
the templates used should be carefully checked and made
publicly available\footnote{ESP templates are available
at the following WWW address: 
http:/boas5.bo.astro.it/$^\sim$cappi/esokp.html}, 
as already done by the LCRS group, to discover and quantify
any systematic difference.
Even if the amplitude of the effect is not large, it is quantitatively
more important that typical zero--point shifts, and
it is significant enough to affect for
example measures of velocity dispersions and galaxy peculiar velocities, or
the interpretation of results for very distant galaxies, as those
which are reported by Cowie et al. (1995) and Steidel et al. (1997).

\section*{ACKNOWLEDGEMENTS}
We thank D.Tucker and H.Lin for useful information. Lin provided us with
the 3 templates used for cross--correlating LCRS spectra.
We thank also the referee, S.A.Shectman, for his very useful comments.

\end{document}